\documentclass{PoS}\usepackage{bm}
\title{Full Salpeter Equation with Confining Interactions: Analytic
Stability Proof}\ShortTitle{Full Salpeter Equation with Confining
Interactions: Analytic Stability Proof}\author{\speaker{Wolfgang
Lucha}\\Institute for High Energy Physics, Austrian Academy of
Sciences, Nikolsdorfergasse 18, A-1050 Vienna, Austria\\E-mail:
\email{Wolfgang.Lucha@oeaw.ac.at}}

\abstract{The most popular 3-dimensional reduction of the
Bethe--Salpeter formalism for the description of bound states
within quantum field theory is the Salpeter equation, found as the
\emph{instantaneous} limit of the Bethe--Salpeter framework if
allowing, \emph{in addition}, for \emph{free} propagation of the
bound-state constituents. Unfortunately, depending on the chosen
Lorentz nature of the Bethe--Salpeter kernel, which encodes all
interactions between the bound-state constituents, supposedly
stable results of Salpeter's equation with \emph{confining}
interactions exhibit (un-)expected instabilities, probably related
to Klein's paradox. Clearly, bound states may be regarded as
stable if, for appropriate interactions, their energy eigenvalues
or, in the center-of-momentum system, their mass eigenvalues
belong to a \emph{real} and \emph{discrete} (part of the)
\emph{spectrum} that is \emph{bounded from below}. Discreteness
means either that eigenvalues and a possible continuous spectrum
are \emph{disjoint} or that the spectrum is
\emph{purely~discrete}. Some general features of the eigenvalue
spectra of any Salpeter equation, common to all solutions, are
well-established: Since Salpeter's equation proves to be of the
same algebraic structure as the random-phase-approximation
equation familiar in nuclear physics all \emph{bound-state masses
squared} are real. Of course, this implies neither that the mass
spectrum itself is necessarily real nor, even in those cases where
it may be shown to be real, that it is also bounded from below.
Direct inspection reveals that, for a large class of sufficiently
reasonable Bethe--Salpeter kernels---which includes all
interactions of phenomenological relevance in QCD---, the spectrum
of mass eigenvalues consists, in the complex bound-state mass
plane, at most of \emph{real} pairs of opposite sign and
\emph{imaginary} points. In general, if the Lorentz structure of a
confining kernel is a mixture of time-component vector~and scalar
stability is achieved only if the Lorentz structure is
predominantly a time-component vector. For time-component vector
\emph{harmonic-oscillator} kernels, a straightforward stability
proof is given.}

\FullConference{8th Conference Quark Confinement and the Hadron
Spectrum\\September 1--6, 2008\\Mainz, Germany}

\begin{document}

Building on experience gained in earlier investigations focused to
the simpler \emph{reduced} Salpeter equation
\cite{Lucha07:HORSE,Lucha07:StabOSS-QCD@Work07} and its
improvement \cite{Lucha05:IBSEWEP,Lucha06:GIBSEEQP-C7} obtained by
inclusion of full (dressed) propagators for the bound-state
constituents \cite{Lucha07:SSSECI-Hadron07,Lucha07:HORSEWEP}, a
rigorous proof of the \emph{full}-Salpeter solutions' stability
for confining (such as harmonic-oscillator) interactions of
\emph{time-component Lorentz-vector} nature is constructed. For
interactions described, in configuration space, by central
potentials of harmonic-oscillator form, $V(\bm{x})=a\,\bm{x}^2,$
$a\ne0,$ the \emph{integral} equation representing the Salpeter
equation simplifies to a system of second-order homogeneous linear
ordinary \emph{differential} equations. Instabilities should
appear first in \emph{pseudoscalar} bound states, and this problem
is most severe and restrictive for massless~constituents. Such
setting allows for a thorough spectral analysis of the problem:
The interaction of the fermionic bound-state constituents is of
time-component Lorentz-vector nature if both their couplings to
some effective Lorentz-scalar potential are represented by the
Dirac matrix $\gamma^0.$ In this case, bound states exist for
$a>0;$ the \emph{squares} of their masses are determined by the
eigenvalues of the product $O\equiv B\,A$ of two positive
self-adjoint (Schr\"odinger) operators $A$ and $B$ on the Hilbert
space $L^2({\mathbb R}^3),$
defined~by$$A\equiv-a\,\Delta+2\,|\bm{x}|=A^\dag\ge0\ ,\qquad
B\equiv-a\,\Delta+2\,|\bm{x}|+\frac{2\,a}{\bm{x}^2}=B^\dag\ge0\
,\qquad\Delta\equiv\bm{\nabla}\cdot\bm{\nabla}\ .$$Using the
(unique) positive self-adjoint square root
$A^{1/2}=(A^{1/2})^\dag\ge0$ of $A,$ the eigenvalue problem for
$O$ is found to be equivalent to that for the \emph{positive
self-adjoint} operator $Q\equiv A^{1/2}\,B\,A^{1/2}=Q^\dag\ge0$
while the \emph{operator inequality} $A\le B,$ clearly satisfied
by $A$ and $B,$ translates into the relation $A^2\le Q.$ Because
its potential, $2\,|\bm{x}|/a,$ is bounded from below and rises to
infinity for $|\bm{x}|\to\infty,$ the operator $A$ and thus also
the operator $A^2$ have \emph{purely discrete} spectrum consisting
of real eigenvalues which rise to infinity. Then, applying the
minimum--maximum principle to $A^2\le Q$ reveals that our
bound-state masses squared and, since $Q\ge0,$ also all the masses
themselves are part of a \emph{real discrete~spectrum}.


\begin{thebibliography}{99}
\bibitem{Lucha07:HORSE}Z.-F.~Li, W.~Lucha, and F.~F.~Sch\"oberl,
\emph{Stability in the instantaneous Bethe--Salpeter formalism:
harmonic-oscillator reduced Salpeter equation},
\emph{Phys.~Rev.~D} {\bf 76} (2007) 125028 [{\tt arXiv:0707.3202
[hep-ph]}].
\bibitem{Lucha07:StabOSS-QCD@Work07}W.~Lucha and F.~F.~Sch\"oberl,
\emph{Stability of Salpeter solutions}, in \emph{QCD@Work 2007 --
International Workshop on Quantum Chromodynamics: Theory and
Experiment}, edited by P.~Colangelo \emph{et al.}, AIP
Conf.~Proc.~(AIP, Melville, New York, 2007), Vol.~{\bf 964},
p.~318 [{\tt arXiv:0707.1440 [hep-ph]}].
\bibitem{Lucha05:IBSEWEP}W.~Lucha and F.~F.~Sch\"oberl,
\emph{Instantaneous Bethe--Salpeter equation with exact
propagators}, \emph{J.~Phys.\ G: Nucl.~Part.~Phys.}~{\bf 31}
(2005) 1133 [{\tt hep-th/0507281}].
\bibitem{Lucha06:GIBSEEQP-C7}W.~Lucha and F.~F.~Sch\"oberl,
\emph{Generalized instantaneous Bethe--Salpeter equation and exact
quark propagators}, in \emph{Quark Confinement and the Hadron
Spectrum VII: 7\/$^{th}$ Conference on Quark Confinement and the
Hadron Spectrum -- QCHS7}, edited by J.~E.~F.~T.~Ribeiro, AIP
Conf.~Proc.~(AIP, Melville, New York, 2007), Vol.~{\bf 892},
p.~524 [{\tt hep-ph/0610016}].
\bibitem{Lucha07:SSSECI-Hadron07}W.~Lucha and F.~F.~Sch\"oberl,
\emph{Stability of the solutions of instantaneous Bethe--Salpeter
equations with confining interactions}, in \emph{XII International
Conference on Hadron Spectroscopy -- Hadron 07}, edited by
L.~Benussi \emph{et al.}, Frascati Phys.~Ser.~(INFN Laboratori
Nazionali di Frascati, 2007), Vol.~{\bf 46}, p.~1539 [{\tt
arXiv:0711.1736~[hep-ph]}].
\bibitem{Lucha07:HORSEWEP}Z.-F. Li, W.~Lucha, and
F.~F.~Sch\"oberl, \emph{Stability in the instantaneous
Bethe--Salpeter formalism: a reduced exact-propagator bound-state
equation with harmonic interaction}, \emph{J.~Phys.~G:
Nucl.~Part.\ Phys.}~{\bf 35} (2008) 115002 [{\tt arXiv:0712.2947
[hep-ph]}].
\newpage\end{thebibliography}
\end{document}